\documentclass{eceasst}

\usepackage[T1]{fontenc} 
\usepackage{orcidlink}

\usepackage[accsupp]{axessibility}
\newcommand{\authorRef}[1]{\texorpdfstring{\autref{#1}}{}}
\newcommand{\authorOrcid}[1]{\texorpdfstring{\thinspace\orcidlink{#1}\thinspace}{}}

\newenvironment{aiuse}{%
  \par\vspace{11pt}%
  \noindent\textbf{Declaration on the use of AI:}\enspace}{%
  \par\vspace{11pt}}

\newenvironment{data}{%
  \par\vspace{11pt}%
  \noindent\textbf{Data availability:}\enspace}{%
  \par\vspace{11pt}}

\usepackage{verbatim}
\usepackage{xcolor}
\usepackage{listings}
\usepackage{float}
\usepackage{booktabs}
\usepackage{tcolorbox}
\usepackage{amsmath}

\newfloat{listing}{tbp}{lol}
\floatname{listing}{Listing}
\definecolor{lstbg}{gray}{0.95}
\lstdefinestyle{codeexample}{
  basicstyle=\ttfamily\tiny,
  breaklines=true,
  breakatwhitespace=false,
  frame=single,
  backgroundcolor=\color{lstbg},
  columns=fullflexible,
  keepspaces=true,
  showstringspaces=false,
}

\title{MetaConfigurator: AI-Assisted RDF Authoring from JSON Data} 
\author{Felix~Neubauer\authorOrcid{0009-0008-5367-2034}\authorRef{1}, Mahdi~Jafarkhani\authorRef{1}, Kenichi~Endo\authorOrcid{0000-0002-4128-6514}\authorRef{2}, Jürgen~Pleiss\authorOrcid{0000-0003-1045-8202}\authorRef{3}, Benjamin~Uekermann\authorOrcid{0000-0002-1314-9969}\authorRef{1}} 
\institute{
\autlabel{1} Institute for Parallel and Distributed Systems, University of Stuttgart,
\autlabel{2} Institute of Polymer Chemistry, University of Stuttgart,
\autlabel{3} Institute of Biochemistry, University of Stuttgart
\\
\texttt{felix.neubauer@ipvs.uni-stuttgart.de}
} 
\abstract{
Scientific workflows increasingly generate structured JSON data that is easy to exchange but difficult to interpret consistently across systems due to lacking semantic interoperability. While JSON Schema ensures structural validation, it provides no native support for Linked Data semantics.

This paper presents an RDF Authoring View extending the open-source JSON Schema editor MetaConfigurator, enabling researchers to transform existing JSON, YAML, or CSV data into RDF using AI-assisted RML mappings, refine triples, execute SPARQL queries, visualize knowledge graphs, and export RDF serializations within a single integrated web interface. This workflow is supported by ontology-aware IRI auto-completion, bidirectional synchronization between JSON-LD text views and RDF triple tables, and AI-assisted SPARQL query generation from natural language hints.

We demonstrate the workflow using laboratory data from metal-organic framework (MOF) synthesis experiments. Protocol data describing reagents, procedure steps, and quantities is converted from JSON to ontology-based JSON-LD via RML mappings. We then refine the semantic representation, query relationships between experimental conditions and outcomes, and explore the resulting knowledge graph interactively. This integrated environment bridges conventional structured data management with Semantic Web technologies while preserving experimental context and lowering technical barriers through AI assistance.
} 

\begin{document}
\maketitle

\section{Introduction}

The Semantic Web promises improved interoperability, explicit semantics, and cross-dataset integration~\cite{berners2001semantic,janowicz2015geospatial,hogan2021knowledgegraphs} with technologies such as Resource Description Framework (RDF)~\cite{manola2004rdf}, but its adoption is hindered by ontology design effort, identifier and vocabulary management difficulties, and the complexity of mapping heterogeneous data into RDF~\cite{janowicz2015geospatial,hogan2021knowledgegraphs}.
For example, computational experiments and simulation workflows produce large volumes of scientific data~\cite{boulakia2017workflows,wilkinson2025fairworkflows,villamar2025metadata} and, in principle, such data can be modeled and validated directly with RDF-based standards~\cite{rdf_concepts2014,shacl_standard}.
In practice, however, researchers still often work with spreadsheets or machine-readable formats such as XML and JSON rather than native RDF graphs~\cite{patel2019bridging, koutkias2019data, Briney2020Foundational, Kochev2020Your}.

Declarative transformation languages such as RDF Mapping Language (RML) make it possible to convert JSON, XML, or CSV data into RDF~\cite{dimou2014rml}.
Yet defining mappings, editing RDF triples, querying resulting graphs, and visualizing them typically requires switching between several separate tools, and many existing systems focus on only one part of this workflow~\cite{protege,webprotege,karma,openrefine,sparql_generate,rdflib,redland}.

To address this gap, we extend MetaConfigurator\footnote{\url{https://metaconfigurator.org}}, an open-source web application for schema-driven structured data editing~\cite{metaconfigurator2024}, with an \textit{RDF Authoring View}.
MetaConfigurator is primarily a JSON Schema editor, which can also generate a web form for data instance editing, source code in different programming languages, and documentation based on a schema.
Furthermore, it supports importing JSON, CSV, YAML and XML data and inferring a schema from them~\cite{metaconfigurator2025datamodels}.
Our extension integrates RML-based transformation from JSON to RDF, direct RDF and JSON-LD \texttt{@context} editing, ontology-aware Internationalized Resource Identifier (IRI) auto-completion, SPARQL querying, interactive knowledge graph visualization, and AI-assisted drafting of SPARQL queries and RML mappings from user hints. In this way, MetaConfigurator evolves from a schema-based data editor into an integrated environment for creating, transforming, querying, and exploring semantically enriched data.

This paper is based on the master thesis of Jafarkhani~\cite{mahdi_master_thesis}.
First, the design and implementation of the developed Semantic Web extensions is shown (\textbf{Section~\ref{sec:design}}), then an application example is presented (\textbf{Section~\ref{sec:example}}) and finally, we provide conclusions (\textbf{Section~\ref{sec:conclusion}}).

\section{Design and Implementation}\label{sec:design}

This section describes the Semantic Web extensions developed for MetaConfigurator.
MetaConfigurator is an interactive tool for importing, visualizing, and editing research data. Users can also create or load JSON Schemas to validate their data and support additional tasks such as code generation, AI-assisted transformations, and schema-guided data mapping.

To support Semantic Web workflows, MetaConfigurator is extended with dedicated RDF authoring capabilities through the addition of a new \textit{RDF Authoring View}.

\subsection{RDF Authoring View}\label{sec:design:rdf-authoring-panel}

The \textit{RDF Authoring View} enables users to inspect, edit, and query RDF data directly within MetaConfigurator. Since RDF data in the tool are represented using JSON-LD (a JSON-based RDF serialization~\cite{jsonld2014}), input data must first be available in this format before RDF-specific functionality can be used.

If the currently loaded data are not valid JSON-LD, the interface informs the user that conversion is required. Users can either provide an existing JSON-LD document directly or convert their current JSON data into JSON-LD using the integrated JSON-to-JSON-LD conversion dialog.

\subsubsection{RML Mapping Dialog}
\label{sec:design:rdf-authoring-panel:rml-mapping-tool}


\begin{figure}[tbp]
  \centering
  \includegraphics[width=.65\linewidth]{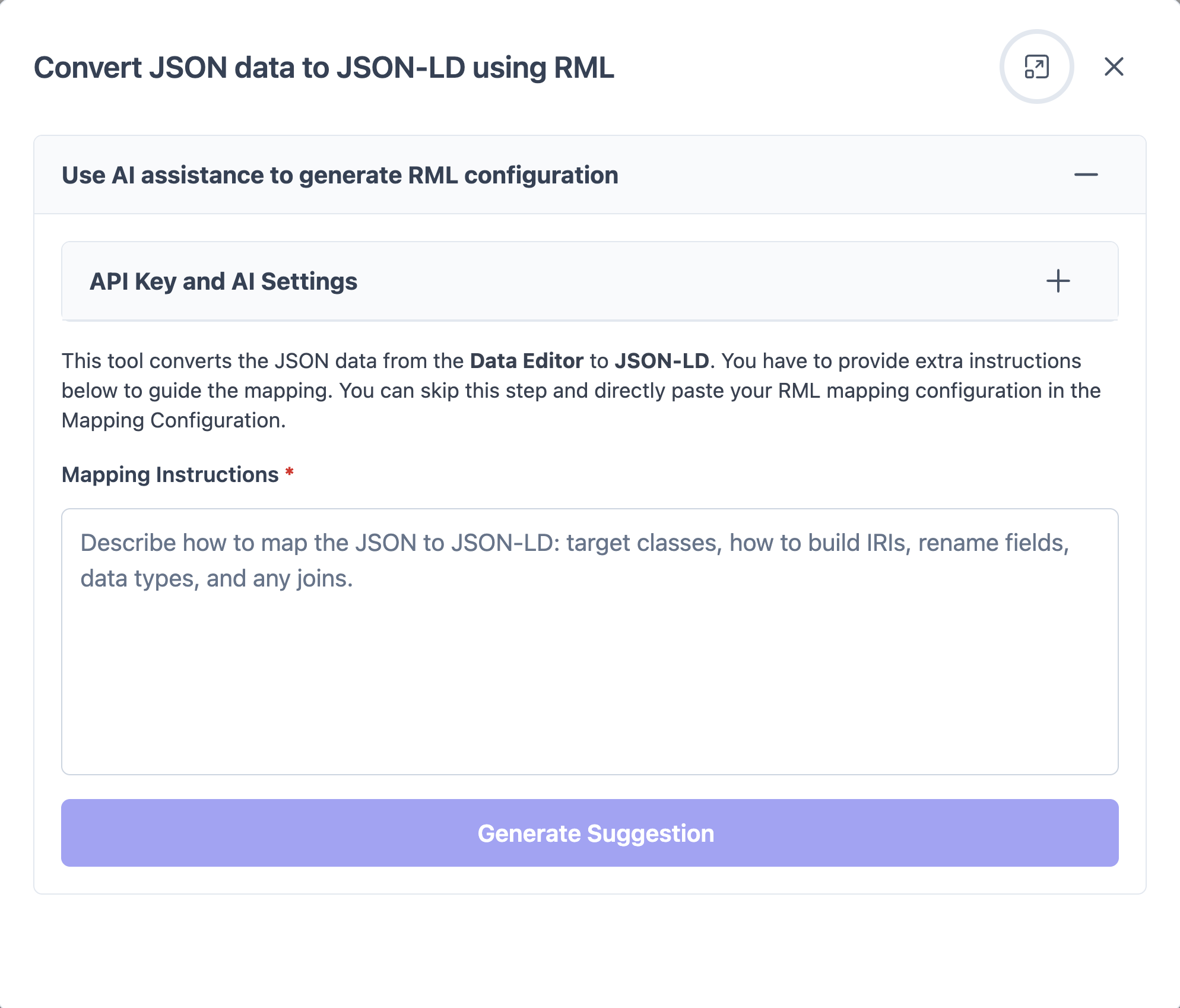}
  \caption{The \textit{RML mapping dialog} for JSON-to-JSON-LD conversion.}
  \label{fig:rdf_authoring_panel:rml_1}
\end{figure}

The dialog (\textbf{Figure~\ref{fig:rdf_authoring_panel:rml_1}}) enables users to convert their existing JSON data into RDF using RML mappings.
Users can either paste an existing RML mapping or generate a mapping suggestion with AI assistance.
For AI-assisted generation, users provide short additional mapping hints in natural language (e.g., which ontologies and terms to use).
These hints are automatically combined with a representative subset of the input data and sent to an external LLM API, together with additional pre-defined instructions.
This follows a similar pattern as the existing AI-assistance functionality in MetaConfigurator~\cite{ai_assisted_schema_mapping_2025}.

The dialog shows the generated or pasted mapping in an interactive code editor with syntax highlighting, which allows inspection or further refinement of the mapping before applying it to the input data.
Once a mapping is finalized, by pressing one button in the dialog the conversion of the JSON data into RDF (serialized as JSON-LD) is performed.

In practice, the creation of suitable RML mappings is usually complex and requires explicit guidance to achieve the intended semantic structure.
Detailed hints become especially important when the target output must follow specific vocabularies, IRI patterns, and inter-entity links.

\subsubsection{View Composition and UI Structure}
\label{sec:design:rdf-authoring-panel:panel-composition}

\begin{figure}[tbp]
  \centering
  \includegraphics[width=.99\linewidth]{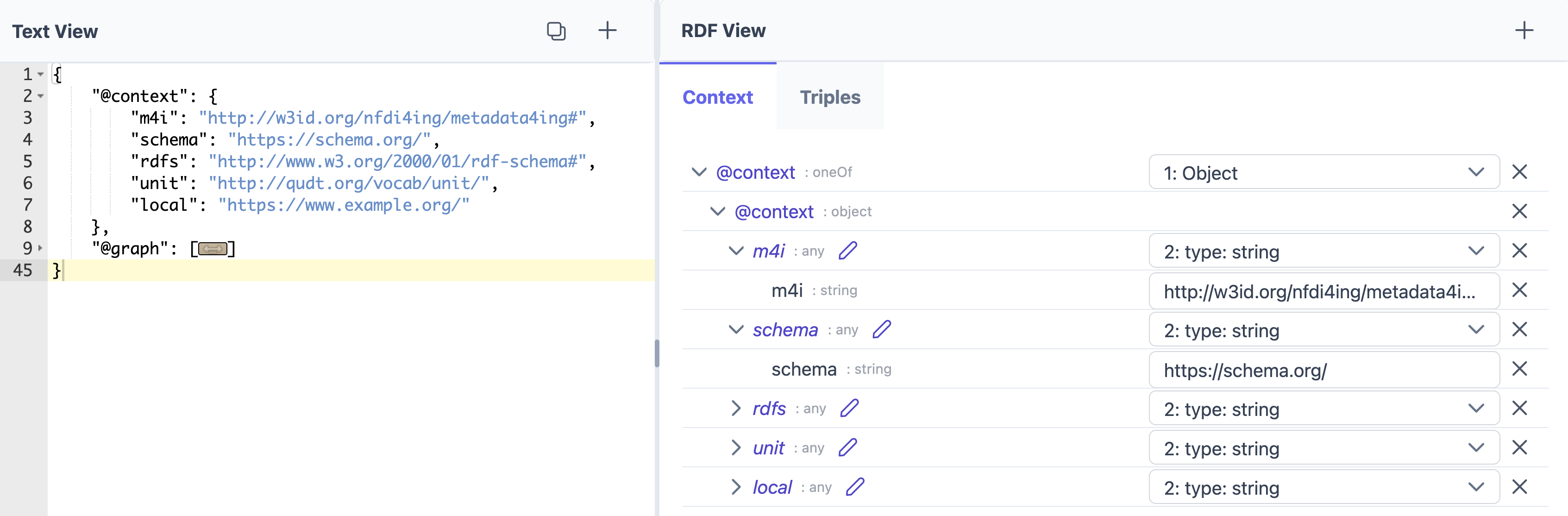}
  \caption{\textit{Context tab} in the \textit{RDF Authoring View}.}
  \label{fig:rdf_authoring_panel:jsonld_context}
\end{figure}

Once the user data is in RDF, the new \textit{RDF Authoring View} can be used for further data inspection, modification and queries.
It is primarily organized into two tabs:

\textbf{Context tab}: for editing JSON-LD \texttt{@context} definitions (\textbf{Figure~\ref{fig:rdf_authoring_panel:jsonld_context}}).

\begin{figure}[tbp]
  \centering
  \includegraphics[width=.99\linewidth]{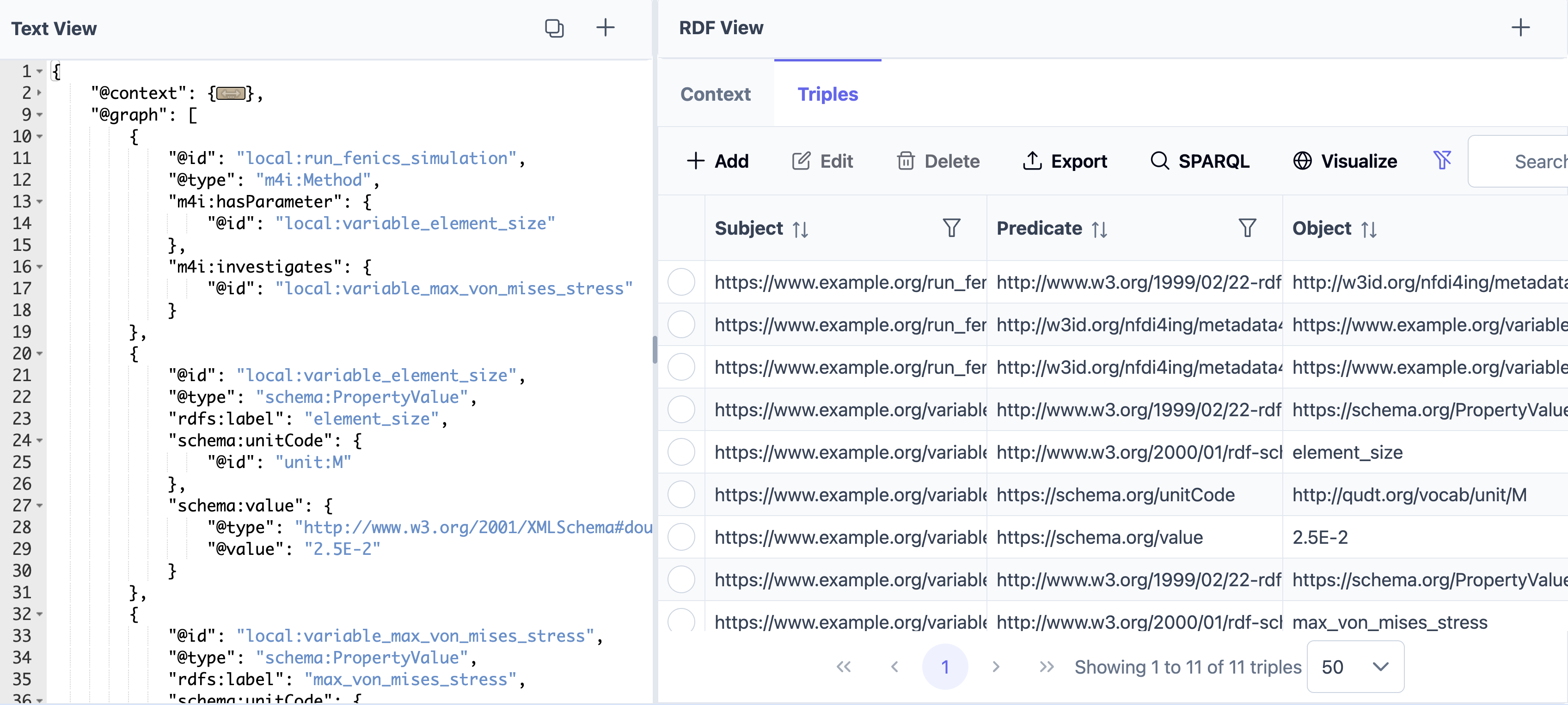}
  \caption{\textit{Triples tab} in the \textit{RDF Authoring View}.}
  \label{fig:rdf_authoring_panel:jsonld_triples}
\end{figure}

\begin{figure}[tbp]
  \centering
  \includegraphics[width=.60\linewidth]{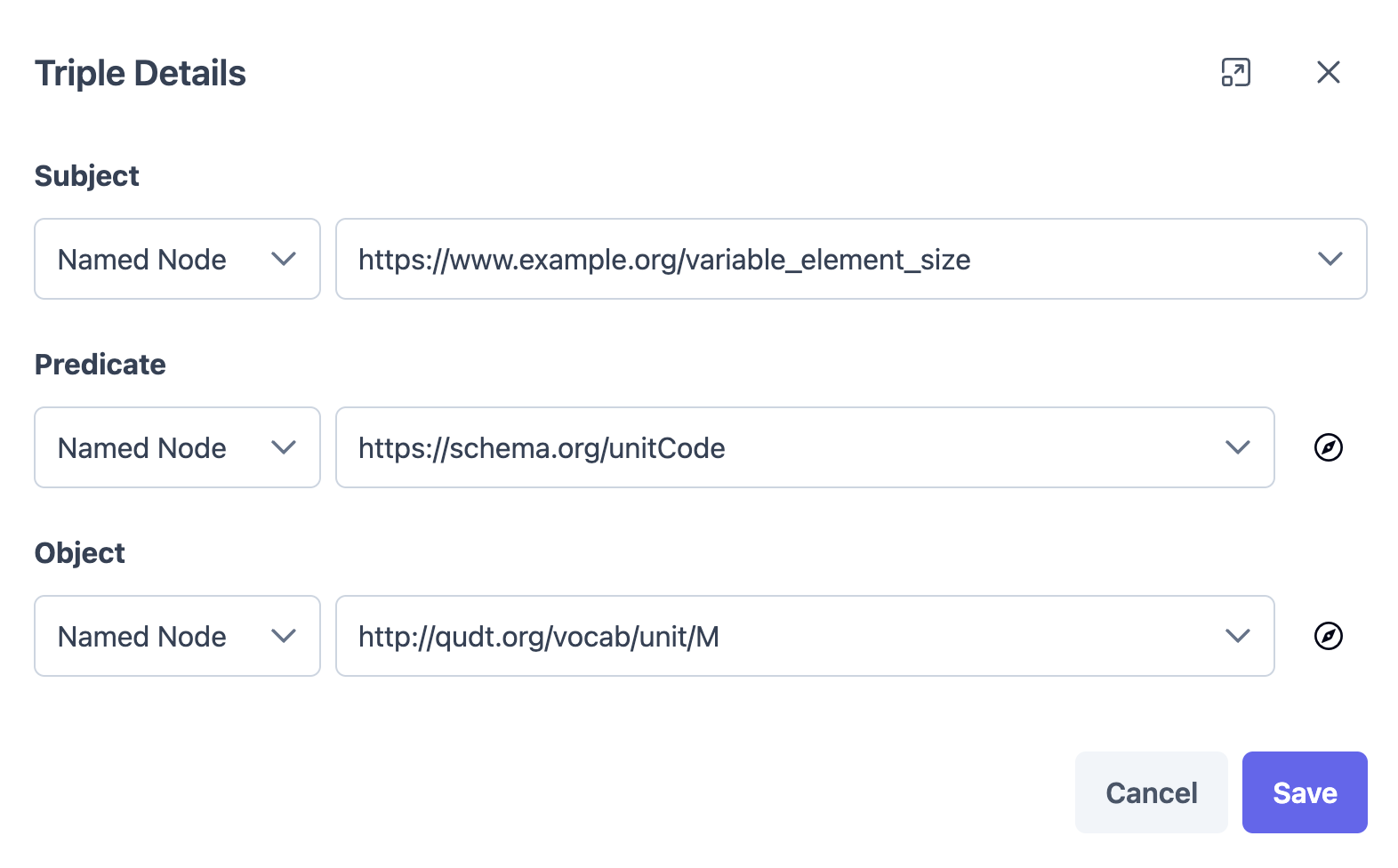}
  \caption{\textit{Edit modal} for triple editing.}
  \label{fig:rdf_authoring_panel:triple_modal}
\end{figure}

\textbf{Triples tab}: for browsing and editing RDF triples.
\label{sec:design:rdf-authoring-panel:triples-tab-implementation}
The \textit{Triples tab} (\textbf{Figure~\ref{fig:rdf_authoring_panel:jsonld_triples}}) shows the RDF triples and supports paging, sorting, and column/global filtering, synchronization of triple selections with other panels, triple modification and export, as well as SPARQL queries, and graph visualization.
\textbf{Figure~\ref{fig:rdf_authoring_panel:triple_modal}} shows the edit modal for triple editing.

\subsubsection{Ontology Integration and IRI Completion}
\label{sec:design:rdf-authoring-panel:ontology-integration}

\begin{figure}[tbp]
  \centering
  \includegraphics[width=.8\linewidth]{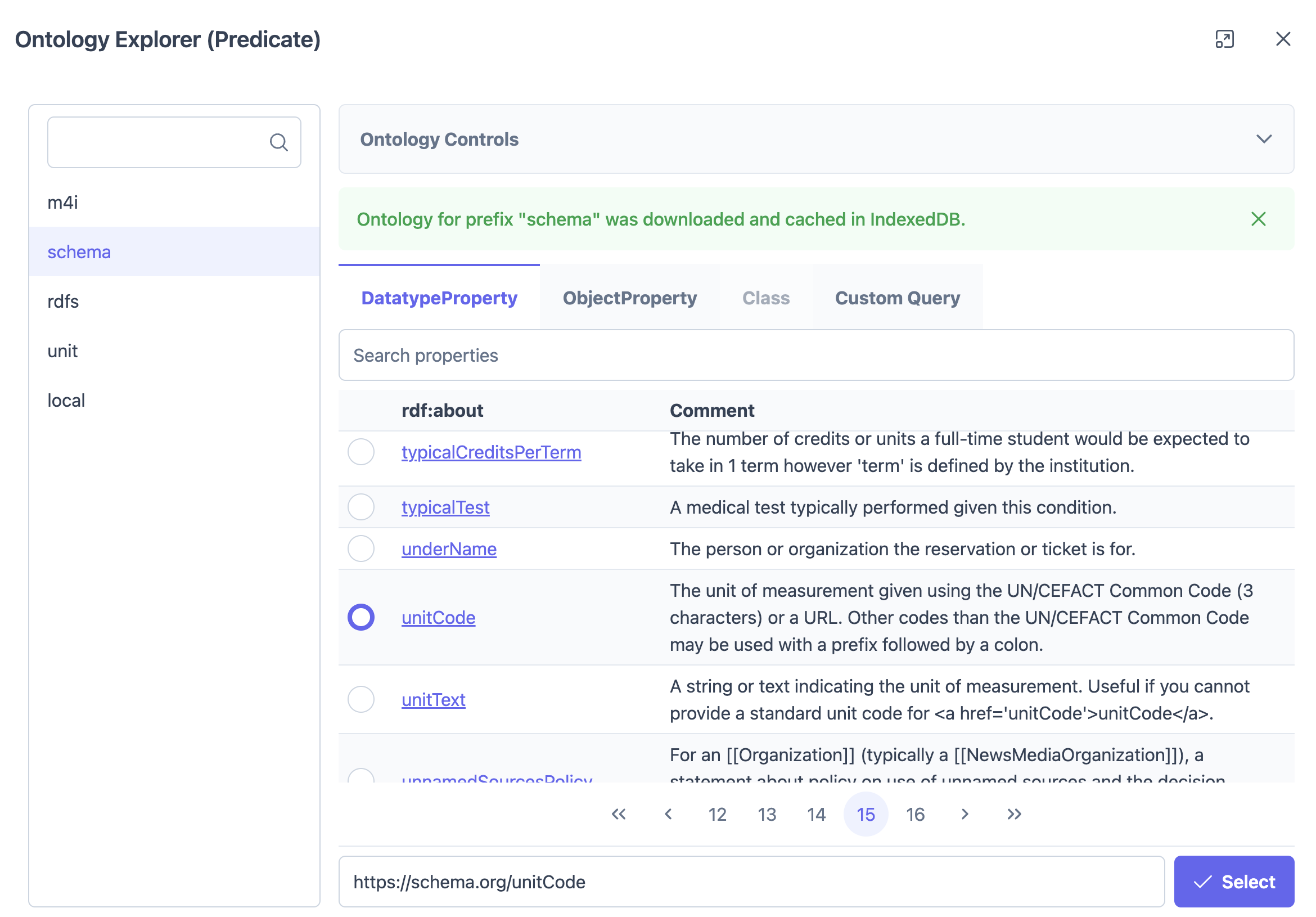}
  \caption{\textit{Ontology Explorer dialog}.}
  \label{fig:rdf_authoring_panel:ontology_integration_2}
\end{figure}

Users can edit IRIs in the \textit{edit modal} (\textbf{Figure~\ref{fig:rdf_authoring_panel:triple_modal}}).
For the \textit{Predicate} and \textit{Object} fields with the \texttt{NamedNode} type, the \textit{RDF Authoring View} provides a completion assistant by opening the \textit{Ontology Explorer dialog} (\textbf{Figure~\ref{fig:rdf_authoring_panel:ontology_integration_2}}).
The explorer reads the JSON-LD \texttt{@context}, extracts prefix mappings, and presents them as selectable ontology scopes.

\subsection{SPARQL Queries}\label{sec:design:query-with-sparql}
SPARQL querying is implemented as a dialog with three tabs: \textit{Query View}, \textit{Result View}, and \textit{Visualization}.
It uses \texttt{Comunica}'s in-browser \texttt{QueryEngine}~\cite{comunica_query_sparql}, which supports SPARQL 1.2 draft query-related specifications according to official documentation~\cite{comunica_supported_specs}.

\begin{figure}[tbp]
  \centering
  \includegraphics[width=.75\linewidth]{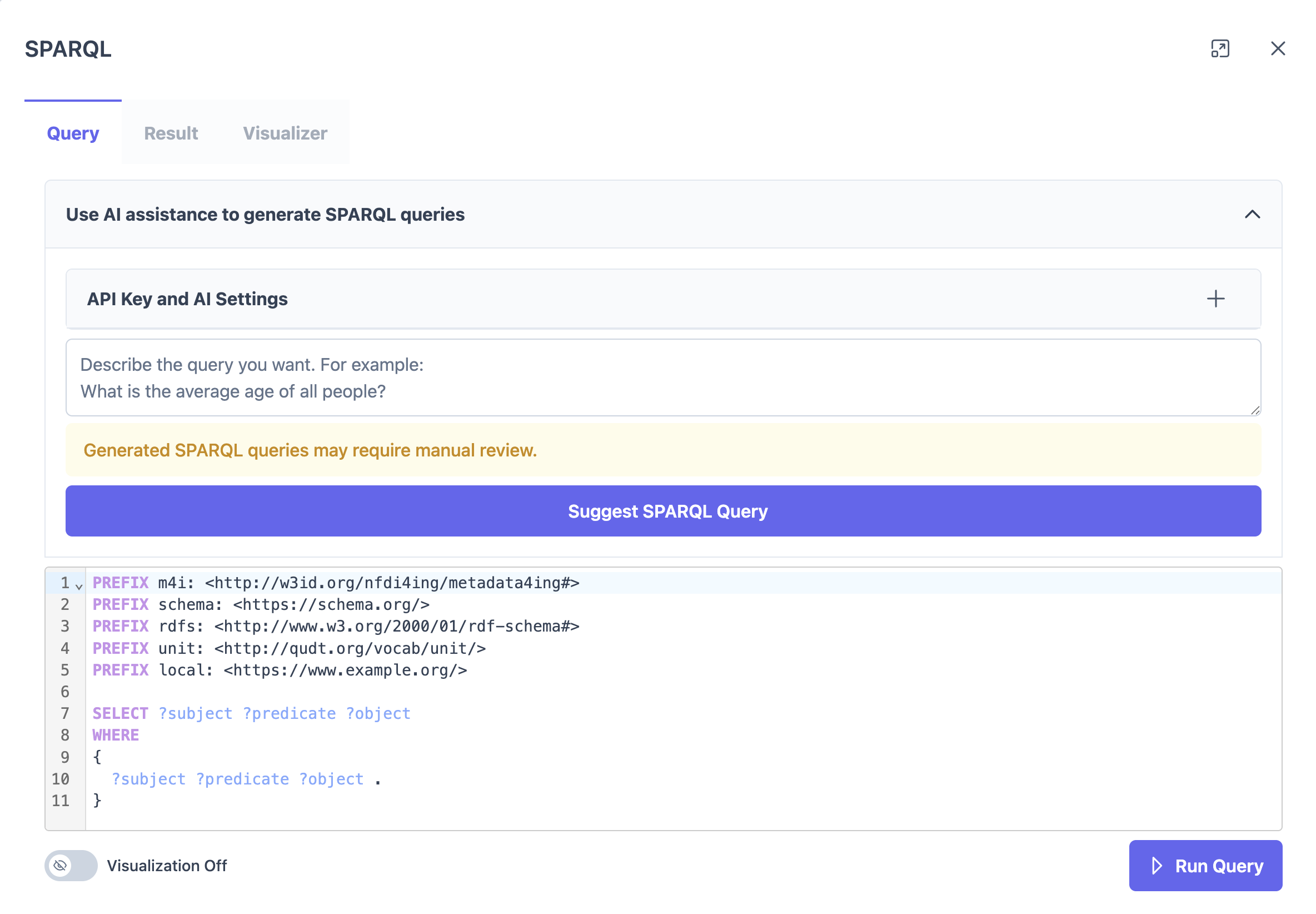}
  \caption{\textit{SPARQL query dialog}, with AI-assisted query generation.}
  \label{fig:rdf_authoring_panel:sparql_ai_generated}
\end{figure}

In the \textit{Query View}, users write and validate queries in an editor and execute them against the in-memory RDF graph.
Syntax checking is performed via debounced live validation using \texttt{Sparql.js}, which provides immediate feedback during editing~\cite{sparqljs}.
The same view also supports AI-assisted query drafting.
As in the RML workflow (\textbf{Section~\ref{sec:design:rdf-authoring-panel:rml-mapping-tool}}), user hints are combined with prompt instructions and representative graph fragments, then sent to the configured AI endpoint (\textbf{Figure~\ref{fig:rdf_authoring_panel:sparql_ai_generated}}).

Human-in-the-loop review is essential in both AI-generated RML mappings and AI-generated SPARQL queries, and users can inspect and refine both results before applying or executing them.





\subsection{Knowledge Graph Visualization and Interaction}\label{sec:design:knowledge-graph-visualization}

Knowledge graph (KG) rendering is implemented using \texttt{Cytoscape.js}~\cite{cytoscapejs}, with the COSE-Bilkent layout extension for force-directed graph placement.
The visualization presents resources as nodes and semantic relations as directed edges, while preserving readable labels through namespace-aware prefix shortening.

The view provides interactive graph navigation features including zoom in/out, fit-to-view, and zoom-to-selected-node actions. For performance and usability, large graphs are detected before rendering and users are prompted to continue or cancel. In addition, graph export is supported as an image, and an optional animation mode can recompute the layout to improve readability.

Node inspection and editing support are integrated.
When a node is selected, a side panel shows the node identifier and its properties, including linkable IRIs. The same panel exposes authoring actions (add/edit/delete property and add/rename/delete node), so users can inspect and refine graph content directly from the visualization workflow.

The \textit{Visualization tab} can be opened either directly from the \textit{Triples tab} or from the \textit{SPARQL query dialog} when the \textit{Visualization On} toggle is enabled.
If opened from the SPARQL dialog, the graph is built from the user's SPARQL CONSTRUCT result and is shown in read-only mode, because returned triples are not guaranteed to map one-to-one to statements in the internal RDF store.

To improve navigation in larger graphs, the visualization also provides node search: users receive suggestions of available nodes while typing, can quickly select the intended node, and the view automatically focuses on it.
The visualization also supports clickable hyperlinks for semantic terms. Predicate labels can open their ontology definitions; for example, \texttt{m4i:hasParameter} links to predicate definition in the ontology\footnote{\url{https://nfdi4ing.pages.rwth-aachen.de/metadata4ing/metadata4ing/index.html\#hasParameter}}.
For object values, \texttt{NamedNode} are handled context-sensitively: if the referenced node exists in the current graph, clicking it focuses that node in the visualization; if it points to an external resource, clicking opens the external definition (e.g., \texttt{unit:M} linking to \url{https://qudt.org/vocab/unit/M}).


\section{Application Example}\label{sec:example}

This section applies the full workflow to a chemical dataset, the synthesis of metal–organic frameworks (MOFs) reported by Neubauer et al.~\cite{neubauer2026data}. MOFs are an emergent class of materials with promising properties~\cite{Canossa_AngewChemIntEd_2025}, but their synthetic procedures include a multitude of critical variables that often cause problems in their documentation and data analysis~\cite{Cheung_AdvancedMaterials_2026}. Recently, JSON and other structured data formats are being developed for machine-readable standardized documentation of their synthetic procedures~\cite{neubauer2026data,Cheung_AdvancedMaterials_2026,Wang_J.Am.Chem.Soc._2026}. Adding links to ontologies and structuring data as RDF can further help semantic embedding of such synthetic data for better interoperability and reusability.

We transform the structured synthesis records modeled in JSON into semantically enriched RDF/JSON-LD with RML, and then use them for ontology-aware querying and graph-based analysis.
The goal is to preserve protocol-level synthesis detail (materials, actions, and conditions) while adding formal semantics so the same data can be interpreted consistently across tools and datasets.

\begin{listing}[tbp]
\begin{lstlisting}[style=codeexample]
{
  "Synthesis": [
    {
      "Hardware": { "Component": [{ "_id": "S-1", "_type": "glass vial" }] },
      "Metadata": { "_description": "S-1", "_product": "MIL-88B (Fe)" },
      "Procedure": {
        "Prep": {
          "Step": [
            { "_vessel": "S-1", "$xml_type": "Add", "_amount": { "Unit": "MilliMOL", "Value": 0.4, "$xml_append": "${Value} ${Unit}" }, "_reagent": "FeCl3", "_order": 0 },
            { "_vessel": "S-1", "$xml_type": "Add", "_amount": { "Unit": "MilliMOL", "Value": 0.4, "$xml_append": "${Value} ${Unit}" }, "_reagent": "Benzene-1,4-dicarboxylic acid", "_order": 1 },
            { "_vessel": "S-1", "$xml_type": "Add", "_amount": { "Unit": "MilliL", "Value": 4, "$xml_append": "${Value} ${Unit}" }, "_reagent": "DMF", "_order": 2 }
          ]
        },
        "Reaction": {
          "Step": [
            { "_vessel": "S-1", "$xml_type": "Sonicate", "_time": { "Value": 30.0, "Unit": "MIN", "$xml_append": "${Value} ${Unit}" }, "_order": 0 },
            { "_comment": "In: oven", "_vessel": "S-1", "$xml_type": "HeatChill", "_temp": { "Unit": "DEG_C", "Value": 120.0, "$xml_append": "${Value} ${Unit}" }, "_time": { "Value": 1, "Unit":"DAY", "$xml_append": "${Value} ${Unit}" }, "_order": 1 }
          ]
        },
        "Workup": {
          "Step": [
            { "_vessel": "S-1", "$xml_type": "WashSolid", "_solvent": "EtOH", "_order": 0 },
            { "_vessel": "S-1", "$xml_type": "Dry", "_temp": { "Unit": "DEG_C", "Value": 120.0, "$xml_append": "${Value} ${Unit}" }, "_time": { "Value": 1, "Unit":"DAY", "$xml_append": "${Value} ${Unit}" }, "_pressure": { "Unit": "PA", "Value": 0, "$xml_append": "${Value} ${Unit}" }, "_order": 1 }
          ]
        }
      },
      "Reagents": {
        "Reagent": [
          { "_id": "FeCl3", "_name": "FeCl3", "_role": "substrate" },
          { "_id": "Benzene-1%252C4-dicarboxylic%2520acid", "_name": "Benzene-1,4-dicarboxylic acid", "_role": "ligand" },
          { "_id": "DMF", "_name": "DMF", "_role": "solvent" },
          { "_id": "EtOH", "_name": "EtOH", "_role": "solvent" }
        ]
      }
    }
  ]
}
\end{lstlisting}
\caption{Representative JSON excerpt from the MOF synthesis data entry \texttt{S-1}}
\label{lst:application_mof_json_example_s_1}
\end{listing}

\textbf{Listing~\ref{lst:application_mof_json_example_s_1}} shows a representative subset of this dataset, encoding one complete synthesis procedure identified as \texttt{S-1}.
Minor adjustments were made to the dataset to support the transformation: for the lists of steps, ordering was explicitly captured in a new field for each step and the \texttt{unit} values were adjusted to be consistent with \texttt{qudt} units.
Using an RML mapping the JSON structures are converted into typed RDF resources.
\textbf{Listing~\ref{lst:application_mof_rml_example_1}} shows a subset of the applied RML mapping.
The complete application example is available on GitHub\footnote{\url{https://github.com/MetaConfigurator/meta-configurator/tree/develop/documentation_user/examples/rdf}}.
In MetaConfigurator, this is the key transition from conventional hierarchical and local JSON data to the Semantic Web.
\textbf{Listing~\ref{lst:application_mof_jsonld_example_s_1}} shows part of the transformed data in JSON-LD format, representing the synthesis run for \texttt{S-1}.

\begin{listing}[tbp]
\begin{lstlisting}[style=codeexample]
@prefix rr: <http://www.w3.org/ns/r2rml#> .
@prefix rml: <http://semweb.mmlab.be/ns/rml#> .
@prefix ql: <http://semweb.mmlab.be/ns/ql#> .
@prefix ex: <http://example.org/> .
@prefix obo: <http://purl.obolibrary.org/obo/> .
@prefix rdfs: <http://www.w3.org/2000/01/rdf-schema#> .
@prefix schema: <https://schema.org/> .

<#SynthesisSource> rml:source "Data.json" ; rml:referenceFormulation ql:JSONPath ; rml:iterator "$.Synthesis[*]" .

<#SynthesisTriplesMap>
  rml:logicalSource <#SynthesisSource> ;
  rr:subjectMap [ rr:template "http://example.org/synthesis/{Hardware.Component[0]._id}" ; rr:class obo:COB_0000082 ] ;
  rr:predicateObjectMap [ rr:predicate schema:identifier ; rr:objectMap [ rml:reference "Hardware.Component[0]._id" ] ] ;
  rr:predicateObjectMap [ rr:predicate obo:RO_0000086 ; rr:objectMap [ rml:reference "Hardware.Component[0]._type" ] ] ;
  rr:predicateObjectMap [ rr:predicate rdfs:label ; rr:objectMap [ rml:reference "Metadata._description" ] ] ;
  rr:predicateObjectMap [ rr:predicate obo:OBI_0000299 ; rr:objectMap [ rml:reference "Metadata._product" ] ] .
\end{lstlisting}
\caption{Part of RML mapping focused on synthesis-level hardware and metadata}
\label{lst:application_mof_rml_example_1}
\end{listing}

\begin{listing}[tbp]
\begin{lstlisting}[style=codeexample]
{
  "@context": {
    "ex": "http://example.org/",
    "obo": "http://purl.obolibrary.org/obo/",
    "rdfs": "http://www.w3.org/2000/01/rdf-schema#",
    "schema": "https://schema.org/",
    "xsd": "http://www.w3.org/2001/XMLSchema#",
    "qudt": "http://qudt.org/vocab/unit/"
  },
  "@graph": [
    {
      "@id": "ex:synthesis/S-1",
      "@type": "obo:COB_0000082",
      "schema:identifier": "S-1",
      "obo:RO_0000086": "glass vial",
      "obo:BFO_0000051": [ { "@id": "ex:step/prep/S-1-0" } ],
      "obo:OBI_0000299": "MIL-88B (Fe)",
      "rdfs:label": "S-1"
    },
    {
      "@id": "ex:step/prep/S-1-0",
      "@type": "obo:OBI_0000070",
      "obo:RO_0000056": "Add",
      "obo:OBI_0001937": { "@id": "ex:quantity/prep/S-1-0/amount" },
      "obo:OBI_0000293": "Prep",
      "obo:RO_0000057": { "@id": "ex:reagent/FeCl3" }
    },
    {
      "@id": "ex:quantity/prep/S-1-0/amount",
      "@type": "schema:PropertyValue",
      "rdfs:label": "amount",
      "schema:unitCode": { "@id": "qudt:MilliMOL" },
      "schema:value": { "@type": "xsd:double", "@value": "4.0E-1" }
    },
    {
      "@id": "ex:reagent/FeCl3",
      "@type": "obo:CHEBI_24431",
      "obo:RO_0000087": "substrate",
      "rdfs:label": "FeCl3"
    }
  ]
}
\end{lstlisting}
\caption{Representative JSON-LD excerpt from the MOF synthesis dataset for \texttt{S-1}}
\label{lst:application_mof_jsonld_example_s_1}
\end{listing}

\begin{figure*}[tbp]
\centering

\begin{tcolorbox}[
    colback=gray!5,
    colframe=gray!50,
    width=\linewidth,
    boxrule=0.5pt,
    arc=2mm
]
\textbf{Natural language hint}
\begin{lstlisting}[style=codeexample, basicstyle=\ttfamily\footnotesize, frame=none, backgroundcolor=\color{white}]
Retrieve all Prep steps with their reagents and amounts for each synthesis.
\end{lstlisting}
\end{tcolorbox}

\vspace{0.5em}
{\Large $\Downarrow$}
\vspace{0.5em}

\begin{tcolorbox}[
    colback=gray!5,
    colframe=gray!50,
    width=\linewidth,
    boxrule=0.5pt,
    arc=2mm
]
\textbf{AI-generated SPARQL query}
\begin{lstlisting}[style=codeexample, frame=none, backgroundcolor=\color{white}]
PREFIX ex: <http://example.org/>
PREFIX obo: <http://purl.obolibrary.org/obo/>
PREFIX rdfs: <http://www.w3.org/2000/01/rdf-schema#>
PREFIX schema: <https://schema.org/>

SELECT ?synthesis ?step ?reagentLabel ?amount ?unit
WHERE {
  ?synthesis obo:BFO_0000051 ?step .

  ?step obo:OBI_0000293 "Prep" ;
        obo:RO_0000057 ?reagent ;
        obo:OBI_0001937 ?amountNode .

  ?amountNode schema:value ?amount ;
              schema:unitCode ?unit .

  ?reagent rdfs:label ?reagentLabel .
}
ORDER BY ?synthesis ?step
\end{lstlisting}
\end{tcolorbox}

\vspace{0.5em}
{\Large $\Downarrow$}
\vspace{0.5em}

\includegraphics[width=.99\linewidth]{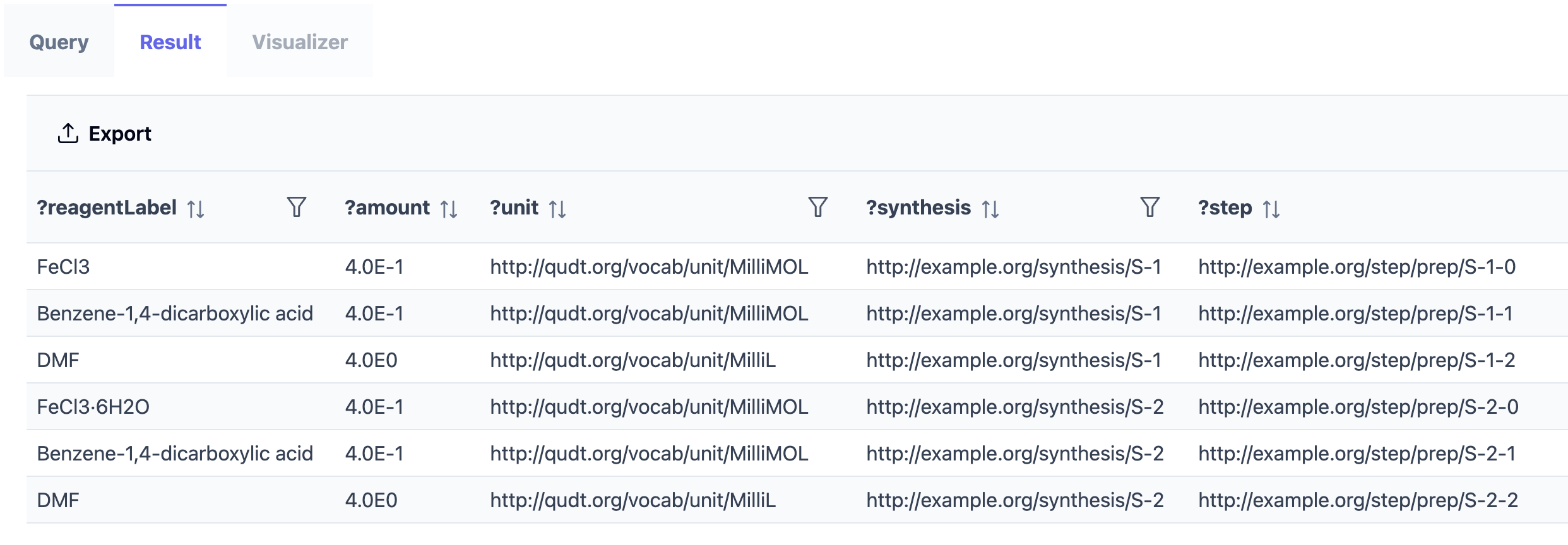}

\caption{AI-assisted SPARQL workflow in the \textit{RDF Authoring View}: a natural language hint is translated into a SPARQL query, which is then executed to produce the query result.}
\label{fig:application:sparql_ai_workflow}
\end{figure*}

After applying the mappings, users can query the resulting RDF graph directly with SPARQL.
In practical laboratory settings, however, writing correct SPARQL queries is often a bottleneck, especially for users with strong domain expertise but limited query-language experience.
To reduce this friction, the AI-assistance can be used.
Users can start from experimental questions instead of manually constructing triple patterns from scratch.
\textbf{Figure~\ref{fig:application:sparql_ai_workflow}} illustrates the AI-assisted SPARQL workflow: a concise natural language user hint is provided to the AI endpoint, which generates a candidate query using a representative subset of the current JSON-LD/RDF graph as contextual input. Executing the generated query retrieves all preparation steps together with their associated reagents and amounts for each synthesis run, as shown in the query result.

This workflow lowers the entry barrier for SPARQL authoring, reduces trial-and-error in the query editor, and lets domain experts focus more on interpretation than syntax details. At the same time, generated queries remain drafts: they may include modeling assumptions or predicate choices that need correction. A user-controlled review step is therefore still required before execution.

\begin{figure}[tbp]
  \centering
  \includegraphics[width=.85\linewidth]{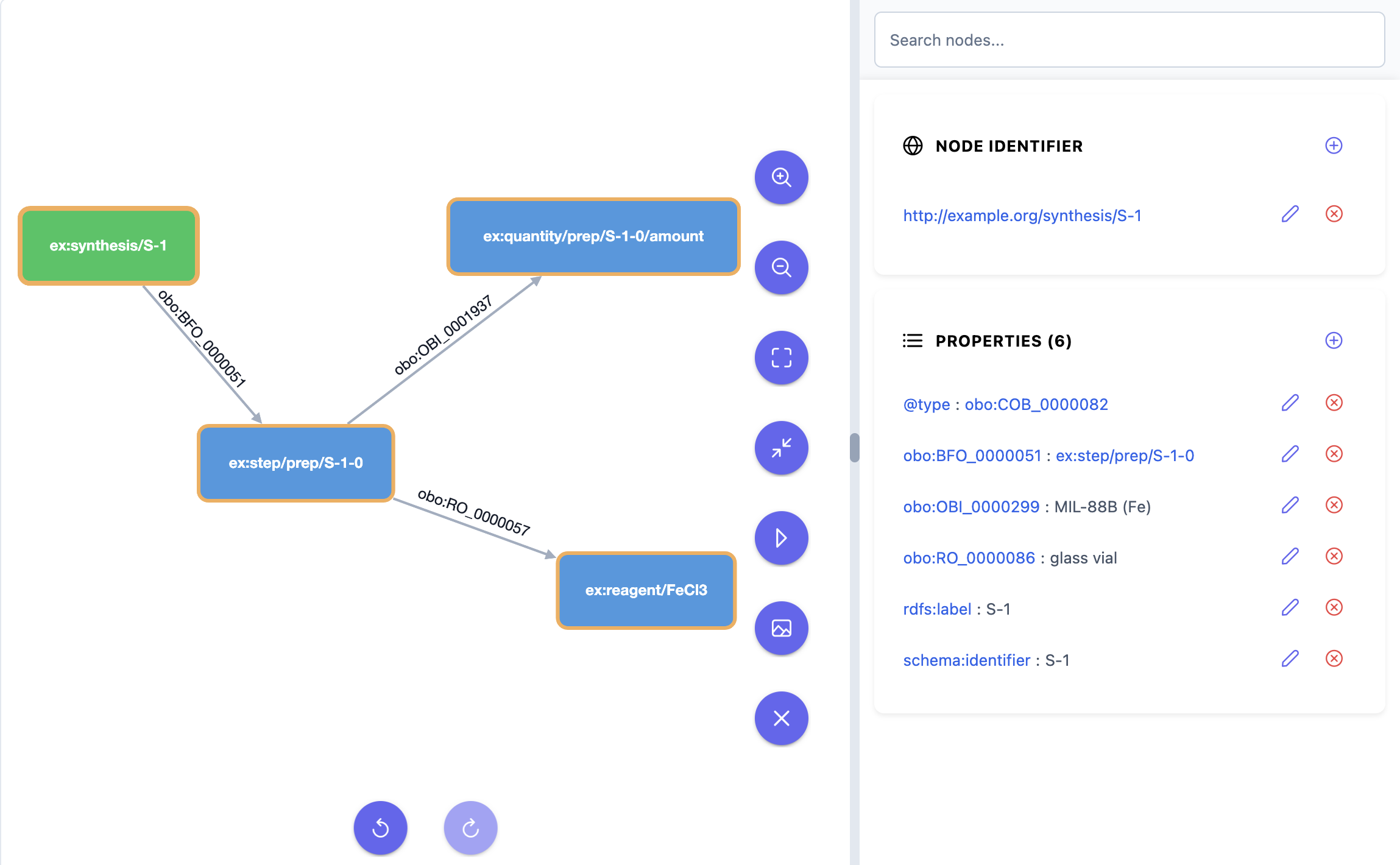}
  \caption{Visualization of the JSON-LD in \textbf{Listing~\ref{lst:application_mof_jsonld_example_s_1}}.}
  \label{fig:application:kg_visualization}
\end{figure}

The visualization view can be opened to inspect the same underlying RDF data as a knowledge graph.
In this representation, synthesis runs, reagents, and process steps appear as connected nodes, while semantic relations are shown as labeled edges. \textbf{Figure~\ref{fig:application:kg_visualization}} shows this view for the mapped MOF synthesis data.

In addition, users can filter the graph semantically with SPARQL and visualize only the query result as a subgraph. This is particularly useful when the full dataset is large: instead of rendering all nodes, users can formulate a focused query (e.g., only Prep-related resources or only selected reagents), execute it, and inspect the returned structure directly in graph form.

\section{Conclusion}\label{sec:conclusion}

This paper presented an extension of MetaConfigurator that connects structured JSON data with Semantic Web technologies in a single user interface.
The implemented \textit{RDF Authoring View} supports a complete workflow from JSON-to-JSON-LD transformation via RML, to semantic editing, SPARQL querying, and interactive KG visualization.
In addition, AI-assisted generation of RML mappings and SPARQL queries lowers the entry barrier for non-expert users while keeping the user in control through explicit review before execution.

The main contribution of this work is a practical, integrated environment for semantic data authoring that combines data transformation, graph editing, query execution, and visualization within one application. The results show that MetaConfigurator can be extended beyond schema-based JSON editing to support ontology-aware RDF workflows without requiring users to switch between multiple tools. This makes the platform a useful bridge between conventional structured data management and Semantic Web-based data integration. The example use case with the chemical dataset shows its readiness for deployment in research data management to increase interoperability and reusability.

At the same time, several limitations remain. AI-generated mappings and queries may still contain structural or semantic errors, so human review remains essential. Especially for creating suitable RML mappings, human expert knowledge still remains indispensable. Scalability is also limited by browser-side processing, especially for large graphs and interactive visualization. Finally, high-quality RDF authoring still depends on selecting appropriate ontologies, classes, predicates, and identifier patterns, which remains challenging for non-expert users.

Overall, the work demonstrates that user-centered semantic authoring can be embedded into an existing structured-data editor and extended step by step toward RDF creation, querying, and graph analysis.
This provides a foundation for future Semantic Web tooling that is more accessible, more integrated, and better aligned with practical scientific workflows.


 \section*{Author Contributions}

 Author contributions are provided using the CRediT (Contributor Roles Taxonomy) standardized contributor roles~\cite{credit}.

 \begin{itemize}
     \item Felix Neubauer: Conceptualization; Methodology; Validation; Supervision; Project administration; Writing – original draft; Writing – review \& editing; Visualization.
     \item Mahdi Jafarkhani: Conceptualization; Methodology; Software; Validation; Investigation; Formal analysis; Writing – original draft; Visualization.
     \item Kenichi Endo: Writing – review \& editing.
     \item Jürgen Pleiss: Supervision; Writing – review \& editing.
     \item Benjamin Uekermann: Supervision; Funding acquisition; Writing – review \& editing.
 \end{itemize}

\begin{acknowledge}
Financial support by the Deutsche Forschungsgemeinschaft (DFG, German Research Foundation) under grant numbers 358283783 (SFB 1333/2 2022), 390740016 (EXC 2075), and 441958208 (NFDI4Chem) is gratefully acknowledged.
\end{acknowledge}

\begin{data}
The developed codes are publicly available on GitHub\footnote{\url{https://github.com/MetaConfigurator/meta-configurator}}.
The complete application example is available on GitHub\footnote{\url{https://github.com/MetaConfigurator/meta-configurator/tree/develop/documentation_user/examples/rdf}}.
\end{data}

\begin{aiuse}
AI-based assistants (including OpenAI GPT and Claude) were used to support code development and limited editorial refinement. All generated outputs were critically reviewed, tested where applicable, and manually revised to ensure correctness and compliance with academic standards.
\end{aiuse}

\bibliographystyle{eceasst}
\bibliography{example}

\end{document}